# Complementarity of which-path information in induced and stimulated coherences via four-wave mixing process from warm Rb atomic ensemble

Danbi Kim[1], Jiho Park[1,2,3], Changhoon Baek[1], Sun Kyung Lee[4, †], and Han Seb Moon[1,2,*]

[1]*Department of Physics, Pusan National University, Geumjeong-Gu, Busan 46241, Republic of Korea*
[2]*Quantum Sensors Research Center, Geumjeong-Gu, Busan 46241, Republic of Korea*
[3]*Quantum Sensors Research Center, Geumjeong-Gu, Busan 46241, Quantum Sensor Research Section, Electronics and Telecommunications Research Institute, Daejeon, 34129, Republic of Korea*
[4]*Quantum Technology Institute, Korea Research Institute of Standards and Science (KRISS), Daejeon 34113, Republic of Korea*
[†]*sklee95@kriss.re.kr*
[*]*hsmoon@pusan.ac.kr*

Complementarity, a fundamental tenet of quantum optics, is indispensable for elucidating the fundamental principles of quantum physics and advancing quantum information processing applications. In the context of wave–particle duality, induced-coherence experiments were understood through the lens of which-path information. Conversely, the stimulated-coherence experiments were explained by using the indistinguishability of the photon statistics of conjugate photons as a means of realizing complementarity. Here, we report a systematic approach for establishing a complementary relationship between the interference visibility, concurrence, and predictability in the crossing of induced and stimulated coherences of two-mode squeezed coherent states. This is achieved using a double-path interferometer involving two independent four-wave mixing (FWM) atomic samples generated via spontaneous and stimulated FWM processes from a warm atomic ensemble of $^{87}$Rb. We demonstrate that the transition from quantum to classical behavior can be characterized by the induced coherence effect, distinguishing between the two-mode squeezed vacuum and coherent states. Moreover, our experimental scheme, employing two FWM atomic ensembles with long-coherent photons, provides valuable insights into the complementarity of which-path information in induced and stimulated coherences.

Induced-coherence experiments have played a crucial role in significantly advancing quantum sensor research, particularly in areas such as quantum imaging and quantum spectroscopy [1-27]. These experiments typically involve experimental schemes based on path-entangled photons, utilizing the induced quantum coherence of two spatially separated biphoton sources generated from distinct nonlinear crystals. In such schemes, the induced coherence arises from the eradication of the which-path information associated with conjugate photons [28-34]. Of particular interest is the intriguing and surprising ability to extract information from undetected photons via correlated single-photon detection, which offers novel perspectives for measurement methods.

These studies raise an important question regarding the transition between a purely quantum-information perspective, which involves the which-way information of photons in a two-path interferometer scheme, and a stimulated-coherence perspective involving the indistinguishability of photon statistics in both conjugate fields within the induced-coherence configuration [32-36]. This question addresses the fundamental nature of induced coherence, bridging the concepts of photon indistinguishability and the stimulated process involving a seed laser. The concept of complementarity in quantum mechanics becomes highly relevant in shedding light on the relationship between classical and quantum physics, as well as the information tradeoff within quantum measurements. However, there have been no reports on the complementary relationship between the distinguishability, predictability, and concurrence of which-way information and the photon statistics of conjugate photons concerning the crossing of induced and stimulated coherences of two-mode squeezed coherent states.

To date, induced-coherence experiments have predominantly relied on spontaneous parametric down-conversion (SPDC) from nonlinear crystals [1-36]. Meanwhile, narrowband spontaneous four-wave mixing (SFWM) photons from atomic ensembles have served as crucial resources for experimental realization in quantum optics, particularly in relation to quantum memory, quantum repeaters, and long-distance quantum networks [37-42]. Notably, bright and resilient SFWM photon pairs generated from hot atomic vapors have been successfully demonstrated in key quantum optics experiments, including Hong–Ou–Mandel (HOM) interference, Franson interference, and entanglement swapping [43-59]. Recently, induced coherence between the signal photons of SFWM photon pairs generated from two distinct hot atomic vapors was demonstrated experimentally [60]. The experimental scheme with two FWM atomic ensembles is an effective method for showcasing the complementarity in the which-path information of induced and stimulated coherences, because of the single-mode narrowband FWM processes from a collective two-photon coherence atomic ensemble.

In this study, we demonstrate the complementary relationship in the transition between induced and stimulated coherences via spontaneous and stimulated FWM processes using two independent warm atomic ensembles. Our investigation focuses on the photon statistics of photon pairs generated during the crossing of induced and stimulated coherences corresponding to the two-mode squeezed vacuum and coherent states. The degree of indistinguishability of idler photons between two interferometry paths upon using coupled Mach–Zehnder interferometer photons can be adjusted using the transmittance amplitude of the signal photons coupled to two warm $^{87}$Rb atomic ensembles. We intend to showcase the complementarity between the visibility of the induced one-photon interference and the distinguishability of photons in the crossing of induced and stimulated coherences, which transition from a two-mode squeezed vacuum state to coherent states, based on the average photon number and transmittance amplitude of the signal photons.

**Experimental schematic**

The proposed experiment involves the use of two separate warm atomic ensemble SFWMs, each generating two-photon coherence between the $5S_{1/2}(F = 2)$ and $5D_{5/2}(F'' = 4)$ states, as shown in Fig. 1(a). This is achieved by employing counter-propagating pump and coupling lasers, as depicted in Fig. 1(b). Counter-propagating signal and idler photon pairs are emitted in the phase-matched direction due to the collective two-photon coherence in the Doppler-broadened cascade-type atomic system [51]. In particular, we note here that an SFWM source obtained from an atomic ensemble offers the advantages of long coherence times and identical atomic ensembles. These unique atomic properties are crucial for induced-coherence experiments that rely on two independent nonlinear media. Consequently, the twin photon pairs generated in the two atomic ensembles of the warm $^{87}$Rb vapor cell are spectrally matched. When the signal photons from Cell1 completely overlap with the signal mode of Cell2, both signal photons from the twin atomic ensembles become indistinguishable.

For induced coherence without a stimulated process, after erasing the which-way information of both idler photons from the two independent hot atomic vapors, both idler photons arrive together at the beam splitter (BS) and interfere in the two coupled interferometers. When the idler photons become indistinguishable upon alignment, we can observe the interference fringe as a function of the path-length difference $\Delta x$ of interference, even though the relative phase relation between the idler photons emitted from two independent atomic vapor cells is not defined.

However, in the proposed system, we can achieve the crossing of the induced and stimulated coherences of two-mode squeezed coherent states by controlling the average photon number of the stimulating seed laser in the signal-photon mode. We investigate the characteristics of the photon statistics of the two-mode squeezed state by measuring the auto-correlation functions between SPD1 and SPD2 according to the average photon number. Under the condition of a two-mode squeezed state realized via FWM processes from two independent warm atomic vapor cells, it is possible to successfully explore the complementary relationship in the transition between the induced and stimulated coherences. Furthermore, we can investigate the complementarity between the visibility of the induced one-photon coherence of the idler photons and the distinguishability of the signal photons with respect to the transmission of the coupled signal photon or seed photon by adjusting the neutral density (ND) filter. Our experimental system makes it possible to effectively demonstrate the complementary relationship between the distinguishability, predictability, and concurrence of the which-way information and photon statistics of the conjugate field. This demonstration is based on both the magnitudes of mode matching and the average photon number of seed photons.

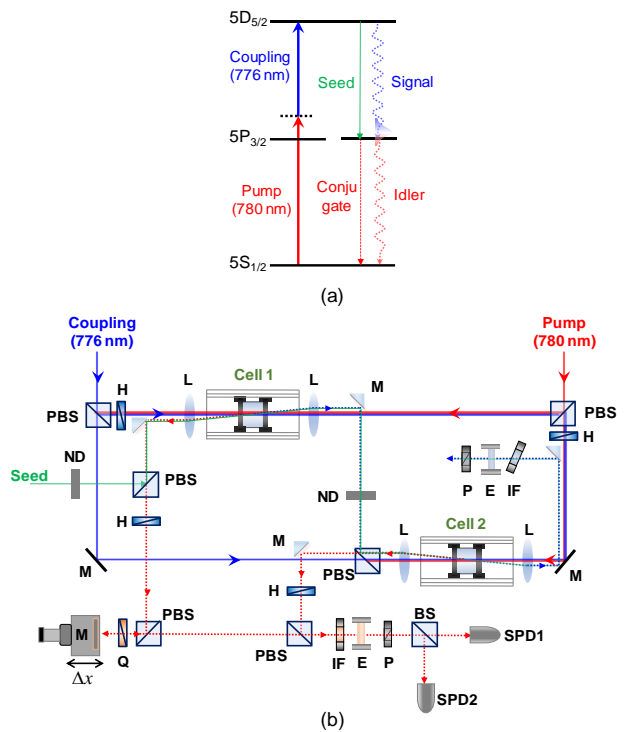

**Figure 1.** Schematic of the experimental setup for investigating the complementary relationship between distinguishability, predictability, and concurrence of the which-way information and photon statistics of the conjugate field using two independent warm $^{87}$Rb atomic vapor cells (M: mirror; BS: beam splitter; PBS: polarizing beam splitter; ND: neutral density filter; H: half-wave plate; Q: quarter-wave plate; E: etalon filter; IF: interference filter; P: polarizer; SPDs: single-photon detectors).

**Photon statistics of two-mode squeezed state** We present the photon statistics of the photon pairs generated from the FWM processes from the warm Rb atomic ensembles, which correspond to a two-mode squeezed coherent state. This is based on the average photon number $|\alpha|^2$ of the photons stimulated by the seed beam in the signal-photon mode, as depicted in Fig. 2. Figure 2(a) illustrates the auto-correlation $G^{(2)}(\tau)$ of the signal photons, which is measured for various $|\alpha|^2$, corresponding to the counting ratio of the stimulated photons to spontaneous photons. Here, $\tau$ denotes the time delay between the detection times of SPD1 and SPD2. For the induced coherence in the case of $|\alpha|^2 = 0$ of Fig. 2(a), $G^{(2)}(\tau)$ exhibits the statistical property of the thermal photons, corresponding to the two-mode squeezed vacuum state. The temporal feature of the $G^{(2)}(\tau)$ curve is independent of the path-length difference $\Delta x$ of the interference arm in the experimental setup displayed in Fig. 1(b). As $|\alpha|^2$ increases, the coincidence counting rates of the background increase; however, the normalized peak values of $G^{(2)}(0)$ significantly decrease.

When $G^{(2)}(\tau)$ is normalized to the coincidence counting rate of the background, we can describe the second-order coherence function $g^{(2)}(\tau)$. As shown in Fig. 2(b), the maximum value of $g^{(2)}(0)$ at $|\alpha|^2 = 0$ is measured to be 1.75 with the SPD timing jitter of ~0.4 ns, which closely agrees with the value of 2 upon adjusting for no SPD timing jitter (See the Supplementary Material for detailed descriptions [61]). Figure 2(b) presents the maximum values of $g^{(2)}(0)$ as a function of $|\alpha|^2$, wherein the left axis (red) represents the measured maximum values of $g^{(2)}(0)$, and the right axis (blue) denotes the reconstructed values obtained by considering no SPD timing jitter. The yellow arrows in Fig. 2(b) correspond to the results shown in Fig. 2(a) for each $|\alpha|^2$ value.

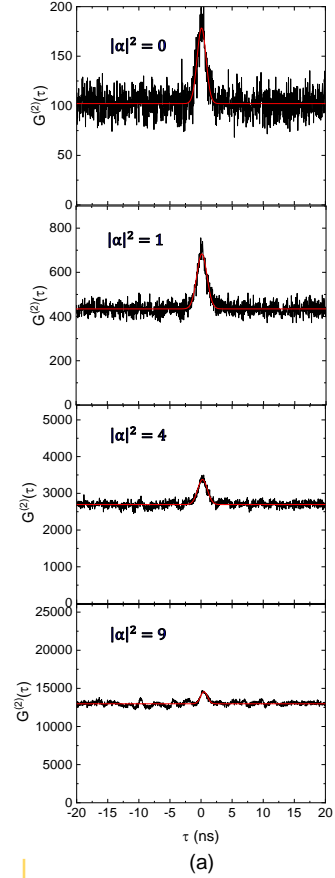

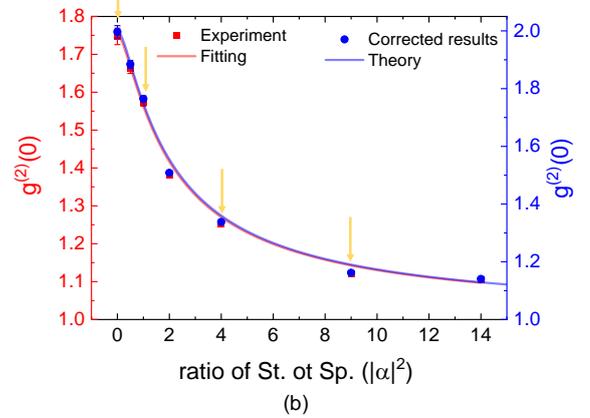

**Figure 2.** Photon statistics of the two-mode squeezed coherent state realized from the FWM processes. (a) Auto-correlation $G^{(2)}(\tau)$ of the signal photons as a function of the counting ratio of stimulated photons to spontaneous photons. (b) Maximum values $g^{(2)}(0)$ of the second-order coherence function of idler photons as a function of $|\alpha^2|$, where the left axis (red) represents the measured results and the right axis (blue) denotes the reconstructed values obtained by considering no SPD timing jitter. The yellow arrows in Fig. 2(b) correspond to the results of Fig. 2(a) for $|\alpha^2|$ = 0, 1, 4, and 9.

The experimental results in Fig. 2(b) can be theoretically explained as corresponding to a two-mode-squeezed coherent state [62]. The squeezing operator is expressed by the Hamiltonian as follows:

$$S(r) = e^{r\hat{a}_i^\dagger \hat{a}_s^\dagger - r^* \hat{a}_i \hat{a}_s}, \quad (1)$$

where $r$ denotes the squeezing parameter. The signal field is initially prepared in a coherent state, $|\psi_0\rangle_{is} = |0\rangle_i |\alpha\rangle_s$, and the maximum values $g_s^{(2)}(0)$ and $g_i^{(2)}(0)$ of the second-order coherence functions of the signal and idler modes, respectively, are expressed as:

$$g_s^{(2)}(0) = \frac{|\alpha|^4 \cosh^4|r| + 4|\alpha|^2 \cosh^2|r|\sinh^2|r| + 2\sinh^4|r|}{(|\alpha|^2 \cosh^2|r| + \sinh^2|r|)^2}, \quad (2)$$

$$g_i^{(2)}(0) = 1 + \frac{2|\alpha|^2 + 1}{(|\alpha|^2 + 1)^2}. \quad (3)$$

In the case of spontaneous FWM, the individual signal and idler modes are thermal fields, that is, $g_s^{(2)}(0) = g_i^{(2)}(0) = 2$. The average photon numbers in the idler and signal modes can be described as $\langle n_i \rangle = (|\alpha|^2 + 1)\sinh^2|r|$ and $\langle n_s \rangle = |\alpha|^2 \cosh^2|r| + \sinh^2|r|$, respectively. As expected, the quantum state of the idler photons becomes coherent when the average photon number $|\alpha|^2$ in the signal mode increases. When $|\alpha|^2 \gg 1$, the $g_i^{(2)}(0)$ value of the idler photons converges to 1 in Eq. (3). The blue curve in Fig. 2(b) represents the theoretical results obtained using Eq. (3) and is independent of the squeezing parameter $r$. The calculated curve is in good agreement with the experimental results. Notably, the transition from quantum to classical features in the induced coherence with increasing $|\alpha|^2$ corresponds to $g_i^{(2)}(0) \to 1$.

**Induced and stimulated coherences** Single-photon interference was measured under the transition condition from a two-mode squeezed vacuum state to coherent states. In coupled Mach–Zehnder interferometry with two embedded FWM atomic ensembles utilizing both spontaneous and stimulated FWM processes, as depicted in Fig. 1(b), the quantum state of photon pairs generated from the two atomic ensembles can be approximated as a path-entangled state. This state is the superposition of the photon pairs created in either the upper or lower atomic ensemble. In the low-induced-photon-number regime, it is well known that there is no one-photon coherence between the signal (idler) photons generated by two independent converters, while two-photon coherence exists between the signal and idler photons.

Figure 3 depicts the single-photon interference measured at SPD1 shown in Fig. 1(b), resulting from the FWMs in the two independent warm atomic ensembles with the two coupled interferometers. The results in Fig. 3(a) show the single-photon interference at $T=1$ for $|\alpha|^2 = 0, 1, 4$, and 9. When the path-length difference $\Delta x$ of the Mach–Zehnder interferometry setup is varied, we observe single-photon interference between the idler photons emitted from the two SFWMs of the upper and lower atomic ensembles (Cell1 and Cell2). For induced coherence at $|\alpha|^2 = 0$, we observed an interference fringe with a visibility of ~0.48. Considering that the induced coherence is not perfect due to experimental limitations, we assume that the offset of the measured interference fringe is the background of ~4,000 counts per 0.2 s. However, as $|\alpha|^2$ increases, the maximum counting number of the constructive interference fringe linearly increases; nevertheless, the minimum counting number is nearly constant, as shown in the cases of $|\alpha|^2 = 0, 1, 4$, and 9 in Fig. 3(a).

In this two-mode squeezed state, the one-photon coherence of two idler photons can be induced by erasing the which-path information of both signal photons. In our experiment, this indistinguishability was achieved either by optimizing the vacuum-field correlation function in the induced coherence by perfectly aligning the two signal modes or by making the photon statistics of the signal photons indistinguishable using a coherent seed laser in the signal mode. From the results shown in Fig. 3(a), we can confirm the crossing of the induced and stimulated coherences of the two-mode squeezed vacuum and coherent states using a double-path interferometer utilizing two independent warm atomic ensembles, generated via spontaneous and stimulated FWM processes.

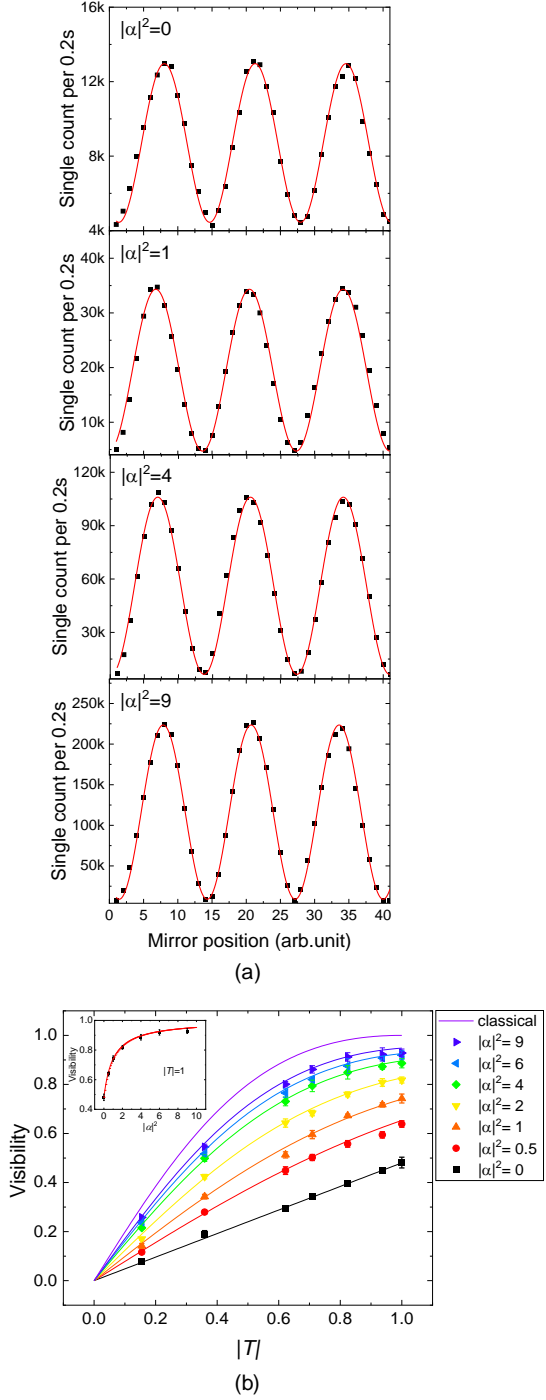

**Figure 3.** Single-photon interferences due to induced and stimulated coherences via both spontaneous and stimulated FWM processes in two independent warm atomic ensembles. (a) Interference fringes as a function of the path-length difference $\Delta x$ for $|\alpha|^2$ = 0, 1, 4, and 9 under the condition of $|T|$ = 1. (b) Visibility of the single-photon interference as a function of the transmittance coefficient $|T|$ for various $|\alpha|^2$ with an inset graph showing the visibility for $|T|$=1. Solid lines represent the theoretical curve for the experimentally measured maximum visibility for $|\alpha|^2$ = 0.

The Hamiltonian of the FWM nonlinear interaction is represented as $H = i\hbar g A \hat{a}_i^\dagger \hat{a}_s^\dagger + h.c$, where $g$ denotes the coupling coefficient and $A$ the joint amplitude of the pump and coupling beams. If an initially coherent signal photon with field amplitude $\alpha$ is stimulated by the seed beam in the signal mode, the initial state can be described as $|\psi(0)\rangle_{is} = |0\rangle_i |\alpha\rangle_s$. If the conversion efficiency and average photon number remain sufficiently low (expressed as $|gAt\alpha| \ll 1$ for an interaction time $t$), then the quantum state of the photon pairs generated from each FWM converter can be approximated as:

$$|\psi(t)\rangle_k = e^{-iHt/\hbar} |\psi(0)\rangle_k$$
$$\approx |0\rangle_{i_k} |\alpha\rangle_{s_k} + gtV|1\rangle_{i_k} \hat{a}_{s_k}^\dagger |\alpha\rangle_{s_k}, \quad (4)$$

where $k \in \{1,2\}$ denotes the upper/lower atomic vapor cells (Cell1 and Cell2), respectively. If the beam splitter is positioned along the signal pathway between the two atomic vapor cells, the mode relation between the two signal paths is given by $\hat{a}_{s_2}^\dagger = T^* \hat{a}_{s_1}^? + R' \hat{a}_{s_0}^?$, where $T$ denotes the transmittance coefficient in the optical path between the two signal photons in the dual-path interferometer. In this case, the quantum state can be expressed as:

$$|\psi(t)\rangle \approx |0\rangle_{i_1} |0\rangle_{i_2} |\alpha\rangle_{s_1} |0\rangle_{s_0}$$
$$+ (c_1 |1\rangle_{i_1} |0\rangle_{i_2} + c_2 T^* |0\rangle_{i_1} |1\rangle_{i_2}) \sqrt{|\alpha|^2+1} |\alpha;1\rangle_{s_1} |0\rangle_{s_0}$$
$$+ c_2 R'^* |0\rangle_{i_1} |1\rangle_{i_2} |\alpha\rangle_{s_1} |1\rangle_{s_0}, \quad (5)$$

where $c_k = g_k A_k t$ ($k \in \{1,2\}$) denotes a nonlinear coupling coefficient, and the single-photon-added coherent state is defined as $|\alpha;1\rangle_s = \hat{a}_s^\dagger / \sqrt{|\alpha|^2+1} |\alpha\rangle_s$ [35, 63]. For identical dual FWM atomic ensembles ($c_1 = c_2$), the single-photon counting rate $R_i$ of the idler photons detected at SPD1 in Fig. 1(b) can be described as:

$$R_i = \langle \psi(t) | \hat{E}_i \hat{E}_i^\dagger | \psi(t) \rangle$$
$$= |gtA|^2 [|\alpha|^2(|T|^2+1) + 2 + 2|T|(|\alpha|^2+1)\cos(\Delta\varphi)], \quad (6)$$

where $\hat{E}_i^\dagger = \hat{a}_{i_1} e^{i\varphi_{i_1}} + \hat{a}_{i_2} e^{i\varphi_{i_2}}$. The phase difference $\Delta\varphi$ is determined by the path differences between the pumps, couplings, and idler photons, as well as the phase difference of the path length of the signal photon passing through an optical sample. Parameter $R_i$ in Eq. (6) represents the summation of the two single-photon counting rates ($R_i^{(I)}$ and $R_i^{(S)}$) of the induced and stimulated coherences, owing to both spontaneous and stimulated processes. In a practical scenario, $R_i^{(I)}$ and $R_i^{(S)}$ can be expressed as:

$$R_i^{(I)} = |gtA|^2 [2 + 2|T|\gamma \cos(\Delta\varphi)], \quad (7)$$

$$R_i^{(S)} = |gtA|^2 |\alpha|^2 [|T|^2 + 1 + 2|T|\cos(\Delta\varphi)], \quad (8)$$

where $\gamma$ denotes the vacuum-field correlation function in the induced coherence [64] and represents the degree of mode overlapping between the two signal photons in the dual-path interferometer. Here, the two corresponding normalized states of all the remaining intrinsic (usually continuous) degrees of freedom (polarization, temporal, spatial, spectral mode, and other factors) of the single photon with correlation correspond to $|\gamma| \leq 1$ [65]. The experimental results in Fig. 3(a) are in good agreement with the calculated interference fringes $R_i = R_i^{(I)} + R_i^{(S)}$ under the condition of $T = 1$.

For identical dual FWM atomic ensembles, the interference visibility for an arbitrary $T$ and $|\alpha|^2$ is described as:

$$V = \frac{2|T|(|\alpha|^2 + \gamma)}{|\alpha|^2(1+|T|^2) + 2} \quad (9)$$

Here, $|T|^2 + |R|^2 = 1$ and $\langle \alpha|\hat{a}\hat{a}^\dagger|\alpha\rangle = |\alpha|^2 + 1$. For the induced coherence case of $|\alpha|^2 = 0$, the visibility is given by $V = \gamma|T|$, where $\gamma = 0.48$ corresponds to the interference visibility at $|\alpha|^2 = 0$, as indicated by the black line in Fig. 3(b). For stimulated coherence in classical regime of $|\alpha|^2 \gg 1$, the visibility as a function of $T$ is given by $V = 2|T|/(1+|T|^2)$, corresponding to the pink curve in Fig. 3(b). To experimentally demonstrate the transition between induced and stimulated coherences via the spontaneous and stimulated FWM processes from two independent atomic ensembles, we measured the interference visibilities as a function of $T$ for the two-mode squeezed coherent states corresponding to $|\alpha|^2 = 0, 1, 4,$ and 9, as shown in Fig. 3(b). The theoretical curves calculated from Eq. (9) are in good agreement with the experimental results. Therefore, our results provide insight into the relationship between the visibility of the induced one-photon interference and the distinguishability of photons in the crossing of induced and stimulated coherences from a two-mode squeezed vacuum state to coherent states.

**Complementarity relations** We investigated the complementarity relation ($K^2 + V^2 = 1$) between the visibility ($V$) of the induced one-photon coherence of idler photons and the distinguishability ($K$) of signal photons with respect to $|\alpha|^2$ and $T$. Distinguishability $K$ can be described as:

$$K = \sqrt{1 - 4|\rho_{12}|^2}, \quad (10)$$

where $\rho_{12}$ denotes the off-diagonal term of the reduced-density matrix of the idler photons, $\rho_i$. For identical nonlinear conversion efficiencies ($c_2 = c_1$) and $|\gamma| = 1$, $\rho_i$, excluding the vacuum state, can be approximated as:

$$\rho_i \simeq |c_1|^2 (|1,0\rangle_i + Te^{-i\phi}|0,1\rangle_i)(_i\langle 1,0| + T^* e^{i\phi}{}_i\langle 0,1|)$$
$$\times \sum_n n|_{s_1}\langle n-1|\alpha\rangle_{s_1}|^2 + |c_1 R'|^2 |0,1\rangle_i \langle 0,1| \sum_n |_{s_1}\langle n|\alpha\rangle_{s_1}|^2$$
$$= |c_1|^2 (|\alpha|^2 + 1)(|1,0\rangle_i + Te^{-i\phi}|0,1\rangle_i)(_i\langle 1,0| + T^* e^{i\phi}{}_i\langle 0,1|)$$
$$+ |c_1|^2 (1-|T|^2)|0,1\rangle_i \langle 0,1|. \quad (11)$$

Considering the normalization condition of $|c_1|^2 = 1/(2 + |\alpha|^2 + |T\alpha|^2)$, the distinguishability of idler photons can be obtained as:

$$K = \left[1 - \left(\frac{2|T|(|\alpha|^2 + 1)}{2 + |\alpha|^2 + |\alpha T|^2}\right)^2\right]^{\frac{1}{2}}. \quad (12)$$

In a bipartite system, duality can be explained by the introduction of concurrence. Distinguishability has a complementary relationship expressed as $K^2 = P^2 + C^2$ between predictability ($P$) and concurrence ($C$). In this study, $P$ refers to the concept of path predictability, defined as a quantitative measure of the a priori which-path knowledge of the two-mode squeezed coherent states via stimulated coherence, and $C$ is defined as a quantitative measure of entanglement. From Eqs. (5) and (11), $P$ and $C$ can be expressed as follows:

$$P^2 = \frac{|\alpha|^4 (1-|T|^2)^2}{(2+|\alpha|^2+|\alpha T|^2)^2}, \quad (13)$$

$$C^2 = \frac{4(1-|T|^2)(1+|\alpha|^2)}{(2+|\alpha|^2+|\alpha T|^2)^2}. \quad (14)$$

Consequently, the three-way complementary identity relationship can be described as $P^2 + C^2 + V^2 = 1$. For the induced coherence at $|\alpha| = 0$, distinguishability is equivalent to concurrence as per $K = \sqrt{1-|T|^2} = C$, while the predictability is zero. This implies that the distinguishability is determined solely by the degree of path entanglement, which indicates its quantum nature. In contrast, in the classical regime of $|\alpha| \gg 1$, distinguishability is equivalent to predictability as per $K = (1-|T|^2)/(1+|T|^2) = P$, with concurrence approaching zero, implying that distinguishability is determined by the imbalance in creation probabilities from the two stimulated FWM atomic ensembles, which is solely dependent on *T*. By varying the average photon number of the injected beam, we can observe different dependencies of *P* and *C*, and consequently, *K* and *V*, with respect to the transmittance (See the Supplementary Material for detailed descriptions [61]), while maintaining the complementarity relation. Figure 4 highlights the normalized visibility (*V*) calculated by subtracting the offset of the measured interference fringe (the background single count in Fig. 3(a)) as a function of $|\alpha|^2$, and *T* from the experimental data depicted in Fig. 3(b). In our proposed system, the transition from quantum to classical features was experimentally demonstrated for the first time using our induced-coherence scheme by increasing the average photon number to manifest the contribution of the stimulated process. In the intermediate regime between quantum and classical transitions, we can observe the derivative of visibility for the intermediate value of *T* by increasing the average photon number (See the Supplementary Material for detailed descriptions [61])[66]. This analysis is based on the quantum mechanical description that visibility $V(T,\alpha)$, predictability (*P*), and concurrence (*C*) as per Eqs. (9), (13), and (14), respectively, reflect the extent to which the quantum entanglement modulation contributes to the modulation of indistinguishability.

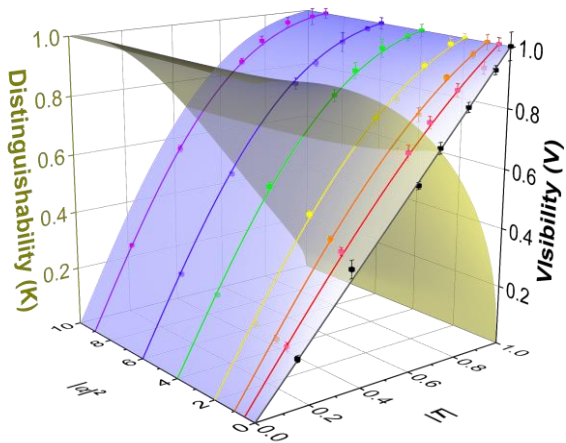

**Figure 4.** Visibility of the one-photon interference as a function of $|\alpha|^2$ and *T*; Complementarity relation ($V^2 + K^2 = 1$) between the interference visibility (*V*) and distinguishability (*K*), where $K^2 = P^2 + C^2$ for predictability (*P*) and concurrence (*C*) in the transition between quantum and classical regimes via the spontaneous and stimulated FWM processes from two independent atomic ensembles.

In conclusion, we experimentally demonstrated the complementary relationship of wave–particle duality in the crossing of quantum and classical regimes using a double-path interferometer that utilized two independent warm atomic ensembles. The quantitative complementarity of wave–particle duality was achieved to measure the interference visibility and tune the distinguishability by adjusting the transmittance of the photons coupled with the double-path interferometer. In particular, we confirmed the transition from quantum to classical features in induced-coherence processes by measuring the photon statistics of photon pairs generated from the FWM processes in warm Rb atomic ensembles, which correspond to a two-mode squeezed coherent state based on the average photon number $|\alpha^2|$ of the stimulated photons. For the first time, we also demonstrated the complementarity relation ($V^2 + C^2 + P^2 = 1$) between the interference visibility (*V*), predictability (P), and concurrence (C), in the intermediate regime between quantum and classical regimes via the spontaneous and stimulated FWM processes from two independent atomic ensembles with long-coherent photons. Our results contribute significantly to the understanding of complementarity in the which-path information of induced and stimulated coherences.

**Supplementary materials**

See the Supplementary Materials for detailed descriptions: limitation of the maximum $g^{(2)}(0)$ value of thermal light from Doppler-broadened warm atomic vapor; theoretical results regarding the calculation of visibility, distinguishability, predictability, and concurrence with respect to the transmission and average photon number in the induced coherence.

**Acknowledgment**

This study was supported by the National Research Foundation of Korea(NRF) grant funded by the Korea government(MSIT) (No. NRF-2021R1A2B5B03002377 and RS-2023-00283146662182065300101), the Institute of Information & Communications Technology Planning & Evaluation (IITP) (No. IITP-2024-2020-0-01606 and 2022-0-01029), and Regional Innovation Strategy (RIS) through the NRF funded by the Ministry of Education (MOE) (2023RIS-007). J.R. acknowledges the National Research Foundation of Korea (NRF) (No. NRF-2020M3E4A1079792 and NRF-2022M3K2A1083890) and Development of quantum-based measurement technologies funded by Korea Research Institute of Standards and Science (KRISS – 2024 – GP2024-0013).

**Conflict of interest**

The authors have no conflicts to disclose.

**Data availability**

The data supporting the findings of this study are available from the corresponding author upon request.